\documentstyle[aps,twocolumn,prl]{revtex}
\def\bea{\begin{eqnarray}}
\def\eea{\end{eqnarray}}
\def\be{\begin{equation}}
\def\ee{\end{equation}}
\def\c#1{\setbox0=\hbox{#1}\ifdim\ht0=1ex\accent24 #1%
  \else{\ooalign{\hidewidth\char24\hidewidth\crcr\unhbox0}}\fi}
\begin{document}
\draft
\author{Krzysztof Sacha and
Jakub Zakrzewski  }
\address{
 Instytut Fizyki imienia Mariana Smoluchowskiego, Uniwersytet
Jagiello\'nski,\\
ulica Reymonta 4, PL-30-059 Krak\'ow, Poland
}
\title{Resonant dynamics of H atom in elliptically polarized
microwave field
}
\date{\today}

\maketitle

\begin{abstract}
The dynamics of Rydberg states of atomic hydrogen driven by elliptically
polarized microwaves of frequency fulfilling 2:1 classical resonance
condition is investigated 
both semiclassically and quantum mechanically in a simplified 
two-dimensional model of an atom. Semiclassical results for quasienergies of 
the system are shown to be 
in a good agreement with exact quantum data.
The structures of the quantum states are found to reflect the 
underlying classical dynamics; especially we show the existence of
nonspreading wavepackets propagating
on elliptical trajectories.  
\end{abstract}
\narrowtext

The pioneering experiment on Rydberg hydrogen (H) atoms ionization 
in microwave fields \cite{bk74} and a subsequent interpretation of
its results in terms of the chaotic classical dynamics \cite{lp79}
 opened up   
a new possibility to study dynamics of atomic systems whose 
classical counterpart reflects a chaotic behavior (for review see
\cite{KoLe95}). Quantum mechanically 
the behavior of atoms in microwave fields is associated with multiphoton processes
what makes the quantum perturbation calculations hardly possible
(many atomic states are strongly coupled)
while, on the other hand, allows to apply semiclassical or even purely
classical description. Signatures of local structures in classical phase
space have been observed in the experimental ionization thresholds where 
an existence of 
stable islands caused increase of the stability of atoms for ionization
\cite{KoLe95}.

The stable structures are 
suitable for a semiclassical quantization. The largest of them are created 
by the primary resonances between the driving field and the unperturbed
Kepler motions of the Rydberg electron. In terms of the principal quantum number
$n_0$ of the unperturbed atomic state, the ratio of the microwave frequency, 
$\omega$, to the Kepler frequency, $\omega_K=1/n_0^3$, has to fulfill
$\omega/\omega_K=s$, where $s$ is an integer number. 
Some of states supported by the resonance islands
reveal interesting properties from the quantum--classical correspondence 
principle point of view. Their time-dynamics may be viewed as a motion of
non-dispersive wavepackets as shown for both circular polarization (CP) 
\cite{ibb} and linear polarization (LP) \cite{ab3} of microwaves. 
These  new states attracted a lot of attention, their decay via
ionization \cite{zdb95} or spontaneous emission \cite{dz98} as well as ways
to populate such states in experiments \cite{dzb95} have been studied.  
The ways of modifying wavepackets properties by additional static fields
were also considered 
 \cite{far_c,szd98}.
Semiclassically, all of them are supported by the principal resonance island, 
i.e. $\omega/\omega_K=1$. The wavepackets built inside a higher primary 
resonance island have been discussed by Holthaus \cite{holt}
in model one dimensional (1D)
systems and to a realistic H atom in LP microwaves
 \cite{bsdz98}. 

Theoretical detection of such interesting states obviously
needs reliable semiclassical methods. In the LP case the 1D
approximation to the dynamics allowed to quantize resonant states exploring
Mathieu equation \cite{sk95}. In the CP case passing to the rotating 
frame removes 
explicit time dependence of the system and the harmonic approximation 
around the stable
fixed point gives the appropriate semiclassical predictions \cite{ibb,dzb95}.
For the realistic LP problem as well as for arbitrary
elliptical polarization (EP) in two dimensional (2D) model
of an atom
the semiclassical method based on the Born--Oppenheimer
approximation allows to describe resonant dynamics \cite{bsdz98,sz98q}. 

This brief report  
extends the previous analysis of the EP case \cite{sz98q}, limited
to the principal resonance island, to a higher primary resonance. 
We describe the full dynamics
of quasienergies as a function of the microwave ellipticity for the 2:1 
resonance case, concentrating in particular on nonspreading wavepackets.
 The experimental data for the EP problem are available 
in the range of the scaled frequency, $\omega_0=\omega/\omega_K$, up to
1.4 \cite{koch96}. Thus a theoretical description of the system for  
$\omega_0=2$ should be
able to be compared with an experiment probably in the immediate future.

As previously \cite{sz98q}, we treat semiclassically
 and quantum mechanically a simplified 2D
model of an atom. Study of such
simplified models have been most successful in the past both for
the LP problem (where 1D model has been a main source
 of quantum results for
a long time \cite{KoLe95}) and for the CP one 
 where also the 2D, polarization plane restricted model has been
utilized \cite{ibb,dzb95} (and references therein).

The Hamiltonian of the hydrogen 2D model atom
driven by an elliptically polarized electromagnetic field reads
in the dipole approximation (in atomic units)
\be
        H=\frac{p_{x}^{2}+p_{y}^{2}}{2}
        -\frac{1}{r}+ F(
        x\cos{\omega t}+\alpha y\sin{\omega t}),
\label{ham}
\ee
where $r= \sqrt{x^2+y^2}$ while $F$ and $\omega$ denote
 the amplitude and the frequency of the microwave
field, respectively.
$\alpha$ defines the ellipticity of the microwaves, with
$\alpha=0$ ($\alpha=1$) corresponding to a LP (CP) limiting case.

Using the Floquet theorem \cite{s65} the solution of the quantum problem
is equivalent to diagonalizing the Floquet Hamiltonian,
${\cal H}=H-i\partial/\partial t$ with time-periodic boundary conditions,
 getting the eigenvalues (quasienergies) and time periodic eigenstates
(Floquet states). The semiclassical quantization of resonant dynamics,
based on a prescription of \cite{holt}, 
closely resembles
the similar procedure applied by us recently for the LP
and EP case \cite{bsdz98,sz98q}. 
The method bases on passing to the extended phase space
by defining the momentum $p_t$ conjugate to the $t$ (time) 
variable what yields the new Hamiltonian, ${\cal H}=H+p_t$ \cite{lil}.
The quasi-energies of the system will be then
 the quantized values of ${\cal H}$.
As the next step we express the Hamiltonian
 in action-angle variables of the unperturbed Coulomb problem \cite{sz97}.
 For the 2D model atom those are e.g.
 the canonically conjugate
pairs $(J,\theta)$ and $(L,\phi)$. $J$
is the principal action
 (corresponding to the principal quantum number, $n_0$).
The conjugate angle, $\theta$, determines the position of the
electron on its elliptic trajectory  and depends linearly 
on time, $\theta\sim\omega_K t$, for an unperturbed atom.
$L$ is the angular
momentum (equal to $L_z$ for the 2D motion in the $x-y$ plane)
 while $\phi$
is the conjugate angle (the angle between the Runge-Lenz vector
and the $x$-axis, i.e. the main axis of the polarization ellipse).

Considering the case of the resonant driving,
 i.e. $\omega_0=s$, we  
apply the secular perturbation theory \cite{lil}
to average over the nonresonant
terms which yields the approximate resonant Hamiltonian of the form
\be
{\cal H}_r=-\frac{1}{2s^2\hat{J}^2}-\omega \hat J+
F\Gamma(L,\phi;\alpha)\cos[\hat\theta-\beta(L,\phi;\alpha)]+\hat p_t
\label{EPr}
\ee
where 
\be
\hat\theta=s\theta-\omega t,\ \hat J=\frac{J}{s},\ \hat p_t=p_t+\omega \hat J.
\label{rot}
\ee
The explicit form of $\Gamma(L,\phi;\alpha)$ and $\beta(L,\phi;\alpha)$
is given by Eq.~(2.17) of \cite{sz97}.

The last stage is to quantize the system using the approximate Hamiltonian, 
Eq.~(\ref{EPr}). 
Trivial quantization of 
 $\hat p_t$,  exploring the time periodicity the system
(note that in the rotating frame, defined by
Eq.~(\ref{rot}), the time period is $\tau=2s\pi/\omega$ \cite{bsdz98}),
 yields additive terms
$k\omega/s$ to quasi-energies, where $k$ is an integer number 
 \cite{holt,bsdz98}. Thus 
the spectrum associated with states localized in the
$s:1$ resonance island repeats itself along the 
energy axis at distances $\omega/s$ \cite{holt,bsdz98}.
The radial motion  in $(\hat J,\hat\theta)$
space is much faster than the angular motion in $(L,\phi)$ space
\cite{lr86,sz97}.
Hence, in the spirit of the Born-Oppenheimer 
approximation, one may first quantize the fast radial motion, keeping 
$L$ and $\phi$ fixed, then pass to the quantization of the slow angular motion.
In fact, because of the specific form of Eq.~(\ref{EPr}), 
the order of the quantizations does not matter, one may quantize first
the slow motion yielding discrete values of $\Gamma$ and then go to 
quantization of the fast motion \cite{bsdz98,sz98q}. We consider
below the $s=2$ resonance as a generic example.

We are interested in strongly localized, wavepacket-like states
lying close to the center of the resonance island in fast 
 $(\hat J,\hat\theta)$ variables. Previously \cite{bsdz98,sz98q} we have
 used pure WKB quantization for that motion which, however, works poorly
 when the island size is small. Such is a case for $s>1$ resonances.
 Thus we improve the procedure and expand the
 principal action to the second order around the center of the resonance
 island defining $I=\hat J-(s^2\omega)^{-1/3}$. Eq.(\ref{EPr}) gives then
  a standard pendulum Hamiltonian with the island size
 given by $\sqrt{F\Gamma}$. 
To ensure a maximal radial localization of the electron we consider then
a ground state of the fast motion by taking appropriate quantum
eigenvalues (as given by Mathieu equation solutions \cite{as}), see similar
treatment for 1D systems \cite{holt,sk95}.

The slow angular motion is determined by constant 
values of $\Gamma(L,\phi;\alpha)$.
 To compare the semiclassical predictions to the exact 
quantum calculations, we consider the $n_0=42$ manifold of our 2D model atom.
For resonant driving we take the microwave frequency to be
$\omega=2\omega_K=2/(n_0+1/2)^3$, i.e. $\omega_0=2$. Note that in the 2D model
the effective principal quantum number is half integer.
Fig.~\ref{qep1} shows values of $\Gamma$ as a function of the scaled angular
momentum, $L_0=L/(n_0+1/2)$, and the $\phi$ angle for two different values 
of the field ellipticity, $\alpha$. 
Semiclassically quantized contours, reflecting slow evolution of the electronic
ellipse, are also presented in the figure. 
Those of them which are localized
around extrema of $\Gamma$ correspond to states with well defined electronic
ellipse. Size of the resonance island in the $(\hat J,\hat\theta)$ space 
depends on value of $\Gamma$, thus, only states localized around the maxima
will show strong radial localization too.
Note that $\Gamma$ is equal to zero for circular orbits, i.e.
for $L_0=\pm 1$. It is obvious because circular motion is purely harmonic
and no primary resonance exists except the $1:1$ case.

Consider level dynamics with a change of the field ellipticity.
Fig.~\ref{levd} shows semiclassical and numerical 
results for quasienergies corresponding to the resonantly 
driven $n_0=42$ manifold as a function of $\alpha$, for the 
scaled field amplitude $F_0=F(n_0+1/2)^4=0.03$.
The nice quantitative agreement between semiclassics and numerics is 
achieved (except in the region of broad avoided 
crossings with other levels -- partners in the crossing are not plotted for
clarity) with significant improvement over earlier approach \cite{sz98q}
for small $\Gamma$ region.

This level dynamics is easy to understand by inspection of 
angular motion change
with $\alpha$, Fig.~\ref{qep1}. Semiclassically, for $\alpha=0$,  
all states are degenerate because of the symmetry
of the $(L,\phi)$ space with respect to the $L_0=0$ axis.
The highest degeneracy exists for librational 
states situated in elliptical islands
around $L_0\approx\pm 0.6$, $\phi=0,\pi$ because  the islands 
are identical.
With a small change of $\alpha$, values of $\Gamma$ 
corresponding to negative $L_0$ 
become smaller while those corresponding to positive $L_0$ become
greater.
Quasienergy levels simply follow the increase or decrease of $\Gamma$,
i.e. the greater value of $\Gamma$ the higher the corresponding quasienergy
level.
Thus, for $\alpha>0$, the degeneracy of many states is removed.
Still for $\alpha<0.17$ there exist three pairs of identical elliptical islands
which support identical semiclassical states, see Fig.~\ref{qep1}. 

With further increase of $\alpha$ the islands situated around 
$L_0\approx -0.6$,
$\phi=0,\pi$ shrink and finally disappear --- they support fewer and fewer 
librational states which during an increase of the field ellipticity vault over 
separatrix and become rotational ones. Also the islands situated, for 
$\alpha=0$,
around $L_0=0$, $\phi=\pi/2,3\pi/2$ shrink with a change of the ellipticity.
Additionally they move towards higher negative values of $L_0$. 
For $\alpha$ close to unity all elliptical islands are too small
to support semiclassical states. Then all states are rotational.

As mentioned before strong localization in both angular and radial motion,
supporting the existence of nonspreading wavepackets is expected around
maxima of $\Gamma$.
Thus one expect nonspreading wavepacket character for states localized 
around $L_0\approx\pm0.6$, $\phi=0,\pi$ for $\alpha<0.17$. For the greater
field ellipticity only the islands around $L_0\approx 0.6$, $\phi=0,\pi$
could support nonspreading wavepackets.
However, single wavepacket propagating along the 2:1 resonance
periodic orbit could not fulfill the periodicity of Floquet states because 
the period of the orbit is twice longer than the microwave period.
So one expects Floquet states being linear combinations of two 
wavepackets shifted in $t$,
i.e. a coordinate variable in the extended phase space, 
by $2\pi/\omega$ which exchange their positions after the microwave period
\cite{holt,bsdz98}.

As a representative of such wavepackets propagating on an elliptical
trajectory we have chosen to plot the states localized around 
$L_0\approx -0.6$, $\phi=0,\pi$ for $\alpha=0.1$. Because of the tunneling 
effect a single quantum eigenstate contains a symmetric or antisymmetric 
combination of two 
semiclassical solutions corresponding to the ellipses with the Runge-Lenz 
vector directed parallel or antiparallel to the $x$-axis. 
Linear combinations of two such eigenstates allow to remove one of the
ellipses. The resulting state is localized on a single ellipse but still
it consists of symmetric combination of two wavepackets 
shifted in microwave phase by $2\pi$ (for
the 2:1 resonance case \cite{holt,bsdz98}). 
To separate a single wavepacket one 
has to
find another state localized on the same ellipse but containing 
antisymmetric combination of the wavepackets. The desired state is prepared 
using eigenstates coming from the similar manifold shifted by $\omega/2$
\cite{holt,bsdz98}.
The resulting wavepacket moving on an elliptical 
trajectory is shown in  Fig.~\ref{packets}.
 This is the single wavepacket rotating
in the opposite direction to the direction of the field vector rotation. 
It propagates along the
periodic orbit supported by the 2:1 resonance, thus, the period of the motion 
is twice longer than the microwave period.

In conclusion we would like to stress that the 
wavepacket presented in Fig.~\ref{packets}
is not an eigenstate of the system.
Tunneling effects between the stable classical islands
will be changing the shape of the packet but, at least 
in the example considered, the
corresponding time scale is of order of a few hundreds of the microwave period.
Another mechanism of its destruction is a slow ionization \cite{zdb95}.

The analysis presented is restricted to the 2D model, its validity for the
real three dimensional atom is an open question. 
Certainly, in the limiting LP case, due to the azimuthal symmetry, the 
wavepackets
appear as doughnuts shaped localized functions moving up and down 
(assuming a vertical polarization of LP microwaves) \cite{bsdz98}. 
For the CP case, on the
other hand, the wavepacket motion was found to remain essentially 2D
\cite{zdb95}. 
The interesting problem how the third dimension affects the dynamics for 
the general EP case is left for future work.

Finally we would like to note that the 2:1 resonance is an example of a
general $s:1$ resonance as the angular motion generated by $\Gamma$
is topologically the same for $s\ge 2$.

We are grateful to Dominique Delande for numerous discussions
and the possibility to use his Lanczos diagonalization routines.
Support of KBN
under project No.~2P302B-00915 is acknowledged.
K. Sacha acknowledges financial support from the Foundation for Polish Science.

\begin{figure}
\caption{Two dimensional hydrogen atom illuminated by resonant,
$\omega_0=2$,
elliptically polarized microwaves. The effective
scaled perturbation
$\Gamma_{0}=\Gamma\omega^{2/3}$ is 
plotted as a function of
the scaled angular momentum, $L_0$, and the angle, $\phi$, between the
Runge-Lenz vector and the main axis of the polarization ellipse -- upper row.
Bottom row shows equipotential curves of the angular part $\Gamma$ of
the Hamiltonian ${\cal H}_r$, Eq.~(\protect{\ref{EPr}}), representing the slow
evolution of the Kepler ellipse. The curves correspond to semiclassical
states originating from the $n_0=42$ hydrogenic manifold.
Columns correspond to the different ellipticity of the microwaves
$\alpha=0.1$ (left) and $\alpha=0.5$ (right).
}
\label{qep1}
\end{figure}
\begin{figure}
\caption{Two dimensional hydrogen atom driven by resonant,
$\omega_0=2$,
elliptically polarized microwaves.
Level dynamics, versus $\alpha$ (i.e. the degree of the field
ellipticity), of the semiclassical quasi-energies [panel (a)] of the
states originating from the $n_0=42$ hydrogenic manifold for $F_0=0.03$ 
 compared with the exact quantum results [panel (b)].
}
\label{levd}
\end{figure}

\begin{figure}
\caption{Wavepacket, being a linear combination of 4 eigenstates of the  
hydrogen atom plus elliptically polarized microwaves system with 
the field amplitude $F_0=0.03$,
 frequency
$\omega_0=2$ for $n_0=42$ and the ellipticity
$\alpha=0.1$.
Temporal evolution is plotted at times $\omega t=0$ (top left),
$\pi/2$ (top center), $\pi$ (top right), $3\pi/2$ (bottom left),
$2\pi$ (bottom right). This wavepacket 
rotates on an elliptical orbit in the opposite 
direction to the rotation of the field vector and 
essentially repeats its periodic motion
with period $4\pi/\omega$. It slowly disperses, either because the 4 building
states are not exactly degenerate (tunneling effect) or because it ionizes.
The size of each box is $\pm4000$ Bohr radii in both $x$ and $y$ directions. 
}
\label{packets}
\end{figure}


\begin{references}
\bibitem{bk74}	  J. E. Bayfield and P. M. Koch , 
	 Phys.\ Rev.\ Lett.\ {\bf 33}, 258 (1974).
\bibitem{lp79} J.~G.~Leopold and I.~C.~Percival, J. Phys. B, {\bf 12}, 709,
			(1979).
\bibitem{KoLe95} P.~M.~Koch and K.~A.~H. van Leeuwen,
                Phys.\ Rep.\  {\bf 255}, 289 (1995).	
			    
\bibitem{ibb}          I. Bialynicki-Birula, M. Kali\'nski, and J. H. Eberly,
 Phys.\ Rev.\ Lett.\ {\bf 73}, 1777 (1994).
 
\bibitem{ab3} D. Delande and A. Buchleitner, Adv.\ At.\
 Mol.\ Opt.\ Phys.\ {\bf 35}, 85 (1994); A. Buchleitner and D. Delande,
 Phys.\ Rev.\ Lett.\ {\bf 75}, 1487 (1995).
\bibitem{zdb95} 		      J. Zakrzewski, D. Delande, and A.
              Buchleitner, Phys.\ Rev.\ Lett.\ {\bf 75}, 4015 (1995);
J. Zakrzewski, D. Delande, and A.
              Buchleitner, Phys.\ Rev.\ E {\bf 57}, 1458 (1998).

\bibitem{dz98}  Z.~Bialynicka-Birula and I. Bialynicki-Birula, Phys. Rev. A,
                        {\bf 56}, 3629 (1997);
			D. Delande and J. Zakrzewski, Phys. Rev. A {\bf 58}, 466
                        (1998);
		K.~Hornberger and A.~Buchleitner, Europhys.\ Lett., 
                        {\bf 41}, 383 (1998).		

\bibitem{dzb95} D. Delande, J. Zakrzewski, and A. Buchleitner,
                 Europhys.\ Lett.\ {\bf 32}, 107 (1995);
		   J. Zakrzewski and D. Delande, J.\ Phys.\ {\bf B 30}, L87
                   (1997).
				
\bibitem{far_c} D.~Farrelly, E.~Lee, and T.~Uzer, Phys. Lett. A {\bf 204},
                        359 (1995);
		A.~Brunello, T.~Uzer, and D.~Farrelly, Phys. Rev. Lett.
                        {\bf 76}, 2874 (1996).
				
\bibitem{szd98} K.~Sacha, J.~Zakrzewski and D.~Delande, Europ. Phys. J. D,
                        {\bf 1}, 231 (1998).
		     
\bibitem{holt}   H.~P.~Breuer and M.~Holthaus, Ann. Phys. {\bf 211}, 249 
		    (1991);
		    M. Holthaus,   Chaos, Solitons and Fractals
                {\bf 5}, 1143 (1995).
	
\bibitem{bsdz98}A.~Buchleitner, K.~Sacha, D.~Delande and J.~Zakrzewski,
               Europ. Phys. J. D {\it submitted}.
		    
\bibitem{sk95} L.~Sirko and P.~M.~Koch, Appl. Phys. {\bf B60}, S195 (1995).

	    
\bibitem{sz98q} K.~Sacha and J.~Zakrzewski, Phys. Rev. A, {\it in press} 
			(1998).				    

\bibitem{koch96} M.~R.~W.~Bellermann, P.~M.~Koch, D.~R.~Mariani, and
                 D.~Richards, Phys.\ Rev.\ Lett.\ {\bf 76}, 892 (1996).
		     
\bibitem{s65} J.~H.~Shirley, Phys. Rev. {\bf 138}, B979 (1965).
	   		     		    
\bibitem{lil} A.~J.~Lichtenberg and M.~A.~Liberman,
{\it Regular and Chaotic Dynamics}, 2nd ed. Springer, New York, 1992.

\bibitem{sz97} K.~Sacha and J.~Zakrzewski,
         Phys.\ Rev.\ A {\bf 56}, 719 (1997).
	   
\bibitem{lr86} J.G.~Leopold and D.~Richards,
                J. Phys. B: Atom.\ Mo.\ Opt.\ Phys.\ {\bf 19},
                 1125 (1986).	 
\bibitem{as} M.~Abramowitz and I.~A.~Stegun (Eds.), {\it Handbook of
               Mathematical Functions.} Dover, New York (1972).
\end{references}
\end{document}